%% file: main.tex
\documentclass[sigplan, nonacm]{acmart}

\usepackage{cleveref}
\crefname{lstlisting}{listing}{listings}
\Crefname{lstlisting}{Listing}{Listings}
\crefname{line}{line}{lines}
\Crefname{line}{Line}{Lines}
\usepackage{mathpartir}
\usepackage{stmaryrd}
\usepackage{upgreek}
\usepackage{tikz}
\usetikzlibrary{calc,positioning,arrows}

\usepackage{pgf-umlsd}
\let\oldsequencediagram\sequencediagram
\let\oldendsequencediagram\endsequencediagram
\renewenvironment{sequencediagram}{%
  \hspace*{-7.8mm}\oldsequencediagram%
}{%
  \oldendsequencediagram\hspace*{-.5mm}%
}

\usepackage{listings}  %
\usepackage{comment}
\usepackage{tabularx}

\definecolor{errorbg}{rgb}{0.95, 0.65, 0.65}
\definecolor{goodybg}{rgb}{0.65, 0.95, 0.65}

\definecolor{keywordcolor}{rgb}{0.7, 0.1, 0.1}
\definecolor{tacticcolor}{rgb}{0.0, 0.1, 0.6}
\definecolor{commentcolor}{rgb}{0.4, 0.4, 0.4}
\definecolor{symbolcolor}{rgb}{0.0, 0.1, 0.6}
\definecolor{sortcolor}{rgb}{0.1, 0.5, 0.1}
\definecolor{attributecolor}{rgb}{0.7, 0.1, 0.1}
\definecolor{codehighlight}{rgb}{0.68, 0.85, 0.9}
\definecolor{RoyalPurple}{RGB}{138,43,226}
\definecolor{RoyalBlue}{RGB}{65,105,225}    %

\lstdefinelanguage{haskell}{
  escapeinside={/-@}{@-/},
  morekeywords=[1]{module,import,where,let,in,if,then,else,case,of,data,type,
                   newtype,class,instance,deriving,do,return},
  keywordstyle=[1]\color{blue},
  morekeywords=[2]{Either,LongList,LargeMatrix,Located,Choreo,IO},
  keywordstyle=[2]\color{blue!50!black},
  morekeywords=[3]{True,False,Left,Right},
  keywordstyle=[3]\color{green!50!black},
  morestring=[b]",
  morestring=[b]`,
  stringstyle=\color{green!50!black},
  sensitive=true,
  comment=[l]{--},
  commentstyle=\color{gray},
  literate=
    {->}{{\color{blue}->}}2
    {<-}{{\color{blue}<-}}2
    {::}{{\color{blue}::}}2,
}

\lstMakeShortInline[language=lean, basicstyle=\ttfamily\normalsize]|

\copyrightyear{2026}
\acmYear{2026}
\setcopyright{cc}
\setcctype{by}
\acmConference[TyDe 2026]{11th Workshop on Type-driven Development}{August 26--27, 2026}{Paris, France}
\acmBooktitle{11th Workshop on Type-driven Development (TyDe 2026), August 26--27, 2026, Paris, France}
\acmDOI{10.1145/3838790.3838794}
\acmISBN{979-8-4007-2864-8/2026/08}

\begin{document}

\title[Choreographic Libraries with Proof-Carrying Located Values]{On Eliminating the Impossible with Dependent Types: Choreographic Libraries with Proof-Carrying Located Values}



\author{Simon Daniel}
\orcid{0009-0003-1636-7253}
\affiliation{%
  \institution{TU Darmstadt}
  \city{Darmstadt}
  \country{Germany}
}
\email{simon.daniel@tu-darmstadt.de}

\author{Timon Böhler}
\orcid{0009-0002-9964-7367}
\affiliation{%
  \institution{TU Darmstadt}
  \city{Darmstadt}
  \country{Germany}
}
\email{timon.boehler@tu-darmstadt.de}

\author{David Richter}
\orcid{0000-0002-8672-0265}
\affiliation{%
  \institution{TU Darmstadt}
  \city{Darmstadt}
  \country{Germany}
}
\email{david.richter@tu-darmstadt.de}

\author{Pascal Weisenburger}
\orcid{0000-0003-1288-1485}
\affiliation{%
  \institution{University of St. Gallen}
  \city{St. Gallen}
  \country{Switzerland}
}
\email{pascal.weisenburger@unisg.ch}

\author{Mira Mezini}
\orcid{0000-0001-6563-7537}
\affiliation{%
  \institution{TU Darmstadt}
  \city{Darmstadt}
  \country{Germany}
}
\affiliation{%
  \institution{hessian.AI}
  \city{Darmstadt}
  \country{Germany}
}
\affiliation{%
  \institution{National Research Center for Applied Cybersecurity ATHENE}
  \city{Darmstadt}
  \country{Germany}
}
\email{mezini@informatik.tu-darmstadt.de}

\input{abstract}

\maketitle

\input{introduction}
\input{total-epp}

\input{dead-branches}

\input{chorlean}
\input{casestudies}

\input{related-work}
\input{conclusion}
\input{acknowledgments.tex}

\bibliographystyle{ACM-Reference-Format}
\bibliography{bib}

\appendix

\end{document}

%% file: abstract.tex
\begin{abstract}
With growing complexity, distributed software systems become increasingly challenging to maintain and reason about.
When implementing a distributed protocol, developers must ensure manually that the different components fit together.
Choreographic programming addresses this challenge by specifying global protocols in a single program and projecting them into communicating processes, so-called endpoints.
Recent choreographic approaches are designed as programming libraries that embed this paradigm into a host language like Haskell or Rust.

In these designs, we observe common cases of \emph{partiality}:
unreachable branches in endpoint projection (EPP) and located-value access can trigger runtime errors or undefined behavior, relying on manual discipline of library maintainers rather than being statically type-checked.
Also, some programs require users to write down \emph{dummy} branches that should not be reachable, for example when branching on sum types.

To close this gap, we use the dependently typed Lean programming language to implement a similar choreographic library.
We show how we are able to move from a partial EPP to a total EPP function, and also eliminate cases of partiality in user-written code with pattern matching on sum types.
ChorLean ensures total EPP and safe value access via proof-carrying located values, passing Lean's totality checker without undefined cases, while supporting the same feature set as libraries like MultiChor.

\end{abstract}

%% file: introduction.tex
\section{Introduction}

As distributed software systems grow more complex, maintaining them and reasoning about their behavior becomes increasingly difficult. When developers implement a distributed protocol, they must manually verify that all components interact correctly. Choreographic programming tackles this challenge by describing global communication protocols in a single specification, which can then be automatically projected into interacting processes.

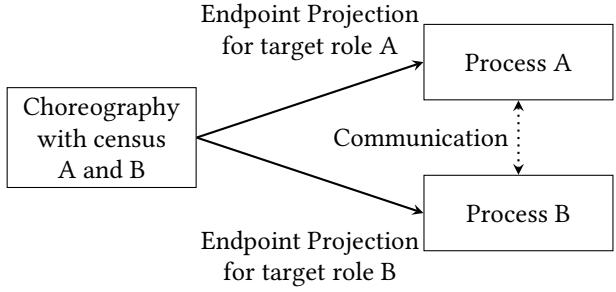
\begin{figure}[h]
\centering
\begin{tikzpicture}[
  node distance=3cm,
  box/.style={draw, rectangle, minimum width=2.5cm, minimum height=1cm},
  arrow/.style={->, >=stealth, thick}
]

\node[box, align=center] (choreo) {Choreography\\with census\\A and B};

\node[box, right=of choreo, yshift=1cm] (prog1) {Process A};
\node[box, right=of choreo, yshift=-1cm] (prog2) {Process B};

\draw[arrow] (choreo.east) -- node[above, pos=0.5, yshift=0.4cm, align=center] {Endpoint Projection\\for target role A} (prog1.west);
\draw[arrow] (choreo.east) -- node[below, pos=0.5, yshift=-0.6cm, align=center] {Endpoint Projection\\for target role B} (prog2.west);

\draw[arrow, <->, dotted] (prog1.south) -- node[left] {Communication} (prog2.north);

\end{tikzpicture}
\caption{Projecting choreographies to processes.}
\label{fig:choreos}
\end{figure}

A choreography describes the global behavior of all roles in a protocol from a global viewpoint.
In particular, choreographies unify sending and receiving into a single communication instruction.
As illustrated in \Cref{fig:choreos}, a programmer writes one choreography (left-hand side) containing code for multiple roles (\emph{A}, \emph{B}, \dots), which endpoint projection (EPP) then maps to one \emph{process} per role (right-hand side), splitting communication into matching pairs of send and receive.
A choreography compiler does this statically, while a choreography library dynamically interprets each operation. The role to which a choreography is projected is called the \emph{target} (sometimes \emph{endpoint}). The set of all roles is its \emph{census}. After projection, processes execute separately---usually on different networked computers---interacting through message passing.

First, choreographic programming was realized via compilers, but more recently, library-based choreographic programming has been put forward.
Where a compiler can generate a static error whenever some part of a protocol cannot be realized,
libraries in most languages do not have the possibility to perform static checks over the total protocol
because they only ever get access to the next choreographic instruction dynamically.
Therefore, checks
that cannot be encoded into the host language's type system are deferred to run time---as done in existing choreographic libraries, such as those in Haskell~\cite{HasChor,MultiChor} or Rust \cite{ChoRus}.
With a sufficiently expressive type system, however, most checks could be delegated to the host language's compiler.

We found that existing libraries still exhibit partiality:
unreachable branches during endpoint projection (EPP) or unsafe value access can cause runtime errors or undefined behavior.
Avoiding these issues typically depends on careful manual oversight rather than type safety.
This shifts the burden of correctness from the type system to informal reasoning and makes it harder to treat endpoint projection as a mathematically total function.
Moreover, users in some systems must explicitly write dummy or undefined branches, especially when handling sum types~\cite{multiply}.

\paragraph*{Contributions.}

In this paper, we present \emph{ChorLean}, our design of a choreographic programming library in the dependently typed Lean programming language.
Our approach provides a \emph{total} implementation of endpoint projection and located-value access, i.e., completely removes partiality.
Safe value access is guaranteed by using proof-carrying located values.

By exploiting dependent types, we design a representation of choreographies and located values in which attempting to access a value without ownership is rejected by the type checker, and endpoint projection is accepted by Lean’s totality checker.
Unlike previous implementations that relied on partiality---evident, for example, wherever \lstinline!undefined! or \lstinline!error! appears in Haskell---our implementation avoids any use of \lstinline!sorry! or \lstinline!panic! (the corresponding Lean terms) in the core EPP and located-value definitions.
Inspired by previous work on verified library-level choreographies in Agda~\cite{Shen2024}, we show that these totality guarantees can be achieved in a practical library without sacrificing expressivity.
Through several case studies, we demonstrate that our design scales to non-trivial choreographies, found in the literature.
We also provide an artifact containing ChorLean's source code \cite{daniel_2026_21531312}.

\paragraph*{Structure of the paper.}
In \Cref{s:total-epp} we illustrate the sources of partiality in existing library-based EPP and how a more expressive host language with dependent types can eliminate them.
In \Cref{s:branches} we illustrate the sources of partiality that appear in user-written code.
\Cref{s:implementation} shows how we encoded choreographies, processes and a total endpoint projection in Lean, and case studies written with our library.
\Cref{s:related-work} discusses related work, and \Cref{s:conclusion} concludes.

%% file: total-epp.tex
\section{Partiality in Libraries}
\label{s:total-epp}

Ideally, the projection from choreographies to processes should be a \emph{total} function, where every choreography maps to a set of interacting processes.
In practice, however, at least three recent libraries~\cite{unwrap-haschor,unwrap-chorus,unwrap-multichor} implement EPP as \emph{partial} functions, with ``impossible'' branches that are not ruled out by types alone.
The consequence is that the compiler cannot help the implementers of choreographic compilers or libraries to avoid these unreachable branches, and instead they fall back to manual discipline, not checked by the compiler.

\begin{figure}
\begin{lstlisting}[language=Haskell, numbers=left, caption={Endpoint projection in HasChor (simplified).}, label={lst:epp-haschor}]
epp :: Choreo m a -> LocTm -> Network m a
epp c l' = interpFreer handler c
  where
    handler :: ChoreoSig m a -> Network m a
    handler (Local l m)
      | toLocTm l == l' =
          wrap <$> run (m (\x -> case x of
            Wrap v -> v
            Empty  -> /-@\colorbox{errorbg}{error "unwrap: Empty"}@-/))/-@\label{epp-1-1}@-/
      | otherwise = return Empty

    handler (Comm s a r)
      | toLocTm s == toLocTm r =
          case a of
            Wrap v -> return (Wrap v)
            Empty  -> return Empty
      | toLocTm s == l' =
          case a of
            Wrap v -> send v (toLocTm r) >> return Empty
            Empty  -> /-@\colorbox{errorbg}{error "send: Empty payload"}@-//-@\label{epp-1-2}@-/
      | toLocTm r == l' =
          wrap <$> recv (toLocTm s)
      | otherwise = return Empty

    handler (Cond l a c)
      | toLocTm l == l' =
          case a of
            Wrap v -> do
              broadcast v
              epp (c v) l'
            Empty  -> /-@\colorbox{errorbg}{error "cond: Empty at broadcaster"}@-//-@\label{epp-1-3}@-/
      | otherwise =
          recv (toLocTm l) >>= \x -> epp (c x) l'
\end{lstlisting}
\end{figure}

Existing libraries represent located values as option-like types and provide an internal “unwrap” operation:
when a role that does not own a value accidentally tries to access it, the unwrap function is undefined~\cite{Shen2024}.
This partiality carries over to endpoint projection, because EPP must manipulate located values internally.

\Cref{lst:epp-haschor} shows (a simplified version of) the endpoint projection function of the HasChor library.
Endpoint projection is the central function in a choreographic library: the correctness of the overall approach is tied directly to the correctness of this function.
Nevertheless, the implementation must resort to runtime errors in several branches.
Lines~\ref{epp-1-1}, \ref{epp-1-2}, and~\ref{epp-1-3} (highlighted in red) use Haskell’s |error| to handle cases in which EPP does not return a value but fails instead, making |epp| a partial function.

The library authors take care to design the surface API so that users cannot easily construct inputs that trigger these errors.
For a well-typed choreography, these branches should be unreachable.
However, this invariant is not expressed in the types of |epp| itself.
At first glance, it is not obvious that the three |error| cases cannot be reached, and nothing in the type system enforces this intention.

\subsection{Avoiding Partiality of Located Value Access}
\label{ss:total}

\providecommand{\coloneqq}{:=}

\begin{figure}
\begin{tabularx}{\textwidth}{@{}>{\centering\arraybackslash}p{0.22\textwidth}@{\hskip 0.00\textwidth}>{\centering\arraybackslash}p{0.22\textwidth}@{}}
\textbf{Partial} & \textbf{Total (ChorLean)} \\[0.5em]
$\alpha \text{\textcolor{red}{@'}O} \coloneqq 1 + \alpha$ &
$\alpha \text{\textcolor{blue}{@}O}\textcolor{blue}{\#}{t} \coloneqq t \in O \to \alpha$ \\[1em]

$\text{un'} : \alpha \text{ \textcolor{red}{@'} O} \to \alpha$ &
$\text{un} : \alpha \text{\textcolor{blue}{@}O}\textcolor{blue}{\#}{t} \to t \in O \to \alpha$ \\[1em]

$\text{un'} \coloneqq
\begin{cases}
  inl~() \mapsto & \colorbox{errorbg}{$\bot$} \\
  inr~v~ \mapsto & v
\end{cases}$
&
$\text{un} \coloneqq id$

\end{tabularx}
\\[1em]
where
  $t$ is the \emph{target role},
  $O$ is a list of \emph{roles},
  $\alpha$ is the \emph{underlying type}.
\caption{Partial vs total located values.}
\label{fig:overview-part-total}
\end{figure}

In general, located values in library-based choreographies have to represent both a value and its absence, depending on whether the current target role executing the program has access to the value.
\Cref{fig:overview-part-total} contrasts the traditional \emph{partial} definition of located values $\alpha \text{\textcolor{red}{@'}O}$, where absence is encoded as $1 + \alpha$ (HasChor, MultiChor, and ChoRus), with our \emph{total} definition $\alpha\text{\textcolor{blue}{@}O}\textcolor{blue}{\#}{t}$, which instead conditions the value on a membership proof $t \in O$.
Located values $\alpha \text{\textcolor{red}{@'}O}$ have an option-like definition that can be either constructed as a |Unit| value, or a value of the underlying datatype |α|.
Additionally, located values are tagged by a subset |O| of roles which we call \emph{owners}.
The intuition behind owners is that located variables carry a value at runtime, if the choreography is projected to a target that is a member of the |owners|.
Roles in ChorLean are identified by the type |Fin N|, which is the type of natural numbers smaller than |N|, where N is the number of roles participating in the choreography.

We could define the |Located'| type corresponding to $\alpha \text{\textcolor{red}{@'}O}$ in Lean as follows:
\begin{lstlisting}
def Located' (O: List (Fin N)) (α: Type)  := Sum Unit α
\end{lstlisting}
|Fin N| is the type of natural numbers smaller than |N| and identifies roles in a choreography.

Then, to access the underlying value of |Located'|,
we have to pattern match on the two possibilities, and provide code for both cases:

\begin{lstlisting}
def unwrap': Located' O α -> α
| .inl _ => /-@\colorbox{errorbg}{sorry}@-/
| .inr v => v
\end{lstlisting}

However, this operation cannot be implemented total in Lean, as there is nothing to return if the located value has been constructed by the |.inl| constructor. To implement the choreography library safely, it should be possible to prove that |unwrap'| is never applied to a value of |.inl ()|.

Existing choreographic libraries solve this problem by providing a safe API on top of option types, which excludes accessing empty values.
Located values can only be unwrapped by the libraries directly, and users get unwrapping functionality in places where they hold some value.
However, internally they still rely on a partial unwrapping function and lack any proof that their implementation of endpoint projection is a total function.
Dependent types in Lean provide an elegant way of closing this gap in the language itself, without relying on tedious manual proofs.

In ChorLean we define the |Located| type, as a total function.
\begin{lstlisting}
def Located 
  (O: List (Fin N)) (t: Hidden (Fin N)) (α: Type) :=
  t.v ∈ O -> α

notation α "@" O "#" t => Located O t α
\end{lstlisting}
The definition of |Located|
adds the projection target |t| as an argument to the type, and maps evidence that |t| \emph{is} a member of the owners to a value of type |α|.
Later we make use of the notation |α @ O # t| for type |α| located at roles |O| with target |t|.

This definition of |Located| enables the intuitive definition of |unwrap| as the identity function, which applies a membership proof |t.v ∈ O| to the located value |Located O t α|:

\begin{lstlisting}
def unwrap: Located O t α -> t.v ∈ O -> α := id
\end{lstlisting}

Such membership proofs can only be constructed, if the target \emph{owns} the value. This provides a \emph{total} implementation of unwrapping a located value.

Throughout our library, we wrap the target role into a special |Hidden| type wrapper, and access the actual target value by |t.v|.
The |Hidden| type is explained in more detail in \Cref{s:implementation}.

\subsection{Choreographic Combinators}

ChorLean provides four primitive constructors of the |Choreo| type (\Cref{ss:basic}), and a small set of derived combinators that cover common patterns of choreographic programming.
\begin{lstlisting}[numbers=left, caption={Derived Choreographic Combinators.}, label={lst-combinators}]
variable {N:Nat} {t: Hidden (Fin N)} {c O: List (Fin N)}
  {p: t.v ∈ c} {α μ:Type} [Serialize μ] 
run      :  {r: Fin N} → {h: t.v ∈ [r]} → IO α  /-@\label{combinators-1}@-/
        → Choreo N t [r] p α
enclave  : (c': List (Fin N)) → c' ⊆ c 
        → ((h : t.v ∈ c') → Choreo N t c' h α)
        → Choreo N t c p (α @ c' # t)
share    : (r : Fin N) → μ @ s::O # t → [s, r] ⊆ c
        → Choreo N t c p (μ @ r::s::O # t)/-@\label{combinators-3}@-/
share'   : (r : Fin N) → (msg: μ @ s::O # t) → [s,r]⊆c/-@\label{combinators-4}@-/
        → Choreo N t c p 
         {lv : μ @ r::s::O # t // ∀ x, msg x = lv ∎}
com      : (r : Fin N) → μ @ s::O # t → [s, r] ⊆ c/-@\label{combinators-5}@-/
        → Choreo N t c p (μ @ [r] # t)
bcast    : μ @ s::O # t → s ∈ c/-@\label{combinators-6}@-/
        → Choreo N t c p μ
bcast'   : (msg: μ @ s::O # t) → s ∈ c/-@\label{combinators-7}@-/
        → Choreo N t c p {v : μ // ∀ x, msg x = v}
locally  : (r : Fin N) → r ∈ c → (t.v = r → IO α)/-@\label{combinators-8}@-/
        → Choreo N t c p (α @ [r] # t)
parallel : ((r : Fin N) → r = t.v → IO α)/-@\label{combinators-9}@-/
        → Choreo N t c p ((r:Fin N) -> α @ [r] # t)
\end{lstlisting}
\Cref{lst-combinators} shows the type signatures of the combinators used throughout the paper.
The |run|, |enclave| and |share| combinators (lines \ref{combinators-1}--\ref{combinators-3}) wrap our primitive constructors |Choreo.Run|, |Choreo.Enclave| and |Choreo.Share|.
|com| (line \ref{combinators-5}) communicates a value like |share|, but moves ownership, such that the result is located only at the receiver.
|bcast| (line \ref{combinators-6}) shares a value with the whole census, returning a plain, un-located value of type |μ|.
The primed variants |share'| (line \ref{combinators-4}) and |bcast'| (line \ref{combinators-7}) attach a proof that the result equals the original located value (see \cref{s:branches}).
|locally| (line \ref{combinators-8}) combines |enclave| and a |run| to execute |IO| at a single role.
|parallel| (line \ref{combinators-9}) runs a program on all roles in the census.

\subsection{A Total Bookseller}

\begin{figure}
\begin{lstlisting}[language=lean, numbers=left, caption={Bookseller Choreography.}, label={lst:capable-books-1}, captionpos=t]
opaque lookup_price: String -> IO Nat 
def B: Fin 2 := 0/-@\label{books-1-0x}@-/
def S: Fin 2 := 1/-@\label{books-1-0}@-/

def books {t p} 
  (budget:Nat): Choreo 2 t [B, S] p Unit :=
  do
  
  let title: String @ [B] <- enclave [B] λ _ => run do/-@\label{books-1-31}@-/
    IO.println "enter your title"
    Input.readString
  let title': String @ [S] <- com S title/-@\label{books-1-7}@-/

  let price: Nat @ [S] <- enclave [S] λ h => run do/-@\label{books-1-3}@-/
    lookup_price (title' h)/-@\label{books-1-4}@-/
  let price' : Nat @ [B, S] <- share B price/-@\label{books-1-41}@-/
  let price'': Nat := price' p/-@\label{books-1-42}@-/
  if budget >= price''/-@\label{books-1-2}@-/
    parallel λ _ _ => IO.println "purchase successful"/-@\label{books-1-8}@-/
\end{lstlisting}%

\begin{centering}
\caption{Bookseller Sequence Diagram.}
\label{books-sequence-simple}
\includegraphics[width=.66\linewidth]{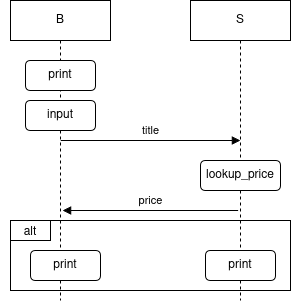}
\end{centering}
\end{figure}

\Cref{lst:capable-books-1} shows a basic bookseller protocol
where a buyer |B| tries to order a book from a seller |S|.
\Cref{books-sequence-simple} outlines the sequence of exchanged messages between both roles.
Using this example, we also illustrate important constructs from ChorLean, such as |run|, |enclave|, |com| and |share|.

The |books| function defines a choreography between roles |B| and |S| (line~\ref{books-1-0x} and \ref{books-1-0}).
Inside, first, |B| prompts the user to enter a book title as a string by using the |enclave| and |run| functions (line~\ref{books-1-31}).
The |enclave| function takes a list of roles, in this case only |B|, and a mapping from a proof that the target is a member of this list, to a choreography with the same census |[B]|.
The |run| function runs a program, but only in a census of exactly one role.

This title is then communicated to |S| using |com S| (line~\ref{books-1-7}), delivering the title variable to the seller's location.
The~\newline |#target| argument in the located type is left out and derived to be |t| from the enclosing choreography.

|S| then looks up the book |price| from a database by\\|lookup_price| (line~\ref{books-1-4}).
To unwrap the value of |price|,
it is shared with |B| using the |share| function and bound to the |price'| variable (line~\ref{books-1-41}).
Instead of transferring the list of owners to |[B]|, it extends the list to |[B, S]|.
From the located variable |price'| we can retrieve a non-located variable |price''| by applying the available evidence |p| (line~\ref{books-1-42}), that |t.v ∈ [S, B]|, such that we can use it in control flow decisions for the choreography.

Next, the choreography checks whether the buyer’s budget is sufficient to cover the price (line~\ref{books-1-2}). If so, both |B| and |S| print the message |"purchase successful"| (line~\ref{books-1-8}). The |parallel| function maps a role, and a proof that this role is the target, to a program that is then run by every role of the census.

In summary, within |run|, IO programs can be executed, such as |lookup_price|.
The application of located variables via a proof extracts their actual value. Evidence of the target being in some list of roles can be gathered by the |enclave| function. Unwrapping located values allows further use in control flow and interaction.

%% file: dead-branches.tex
\section{Avoiding Dead Branches on Sum Types.}
\label{s:branches}

Another case of partiality appears in code that uses library-based choreographies as noted by Bates et al. \cite{multiply}.
Libraries lose information about the value of a communicated message, and thus cannot reuse information gathered about the original value for the result of a communication.

\begin{lstlisting}[caption={Dead branches in MultiChor.}, label={lstdeadbrancheslambda}, language=haskell, numbers=left, float]
sample ::
  Located '["frida"] (Either LongList LargeMatrix) ->
  Choreo ["frida", "emil"] IO ()
sample val = do
  guard <- frida `run` \un ->
    return (isLeft (un frida val)) /-@\label{dead-1}@-/
  broadcast (frida, guard) >>= \case
    True -> do
      list <- frida `run` \un ->
        return (case (un frida val) of 
          Left is -> is
          Right _ -> /-@\colorbox{errorbg}{undefined}@-/)/-@\label{dead-4}@-/
      handleList list/-@\label{dead-2}@-/
    False -> do 
      matrix <- frida `run` \un ->
        return (case (un frida val) of 
          Left _ -> /-@\colorbox{errorbg}{undefined}@-//-@\label{dead-5}@-/
          Right ss -> ss)
      handleMatrix matrix/-@\label{dead-3}@-/
\end{lstlisting}

Consider the choreography written in MultiChor between |frida| and |emil| shown in \Cref{lstdeadbrancheslambda} to see why that is problematic.
The input argument |val| is either constructed as a |LongList| or a |LongMatrix| at |frida|.
Depending on which of both constructors was used (|inl| or |inr|), the input shall be handled by either the |handleList| (line \ref{dead-2}) or |handleMatrix| (line \ref{dead-3}) choreography.
To synchronously make the decision, which follow-up protocol to choose, |emil| has to be informed of the choice.
Exchanging the complete value of |Either LongList LargeMatrix| might not be efficient, or could violate privacy concerns in other applications.
For the branching of the protocol, broadcasting a boolean flag that specifies whether |inl| or |inr| was used would have been sufficient (line \ref{dead-1}).
Retrieval of the underlying value by a consecutive pattern match of the sum (Haskell's Either type in Lean) results in partiality as seen in the redundant required branches in \Cref{lstdeadbrancheslambda} line \ref{dead-4} and \ref{dead-5} (highlighted in red).
Even though developers know that these cases are not reachable, the type-system requires them to either provide a value of some default, or marking it as |undefined| in Haskell.

The dependent type system of Lean is expressive enough to eliminate the need for these branches, making these kinds of choreographies total again.
Even though there is no information about the value of a message sent over the physical network, we can exploit the fact that communicating a value does not change its value for all owning roles during EPP of the |Share| case (\Cref{s:implementation}).

\begin{lstlisting}[numbers=left, caption={Proof Carrying Broadcast.}, label={lst-bcast'}]
variable {N:Nat} {t: Hidden (Fin N)} {c O: List (Fin N)} {p: t.v ∈ c} {μ:Type} [Serialize μ] 

def bcast' {s: Fin N}
  (msg: μ @ s::O # t)/-@\label{bcast'-2}@-/
  (ex: s ∈ c)/-@\label{bcast'-3}@-/
  :
  Choreo N t c p {v: μ // ∀ x, msg x = v} :=/-@\label{bcast'-1}@-/
  multicast' s O c msg ∎ >>= fun v => pure ⟨v.val ∎, ∎⟩
where
  multicast' (s: Fin N)
    (O rs: List (Fin N))
    (msg : μ @ s::O # t)
    (_ex: s::rs ⊆ c) :
    Choreo N t c p
      {v: μ @ rs++s::O # t // ∀ x, msg x = v ∎} :=
    match rs with
    | [] =>
      return ⟨fun x => msg ∎, ∎⟩
    | r::rs => do
      let msg' <- share' r msg
      let temp <- multicast' s (r::O) rs
        (fun _ => msg'.val ∎) ∎
      return ⟨fun x => temp.val ∎, ∎⟩

\end{lstlisting}
By carrying these proofs along communication, we can define a function |bcast'| in \cref{lst-bcast'} that attaches a proof to the resulting value |v|, such that |∀ x, msg x = v| (line \ref{bcast'-1}).
Here, |msg : μ @ s::O # t| (line \ref{bcast'-2}) is the located message of type |μ| with owners |s::O| and target |t|.
|ex| of type |s ∈ c| is the proof that the sender |s| is among the census (line \ref{bcast'-3}).
The result type of |{v: μ // ∀ x, msg x = v}| is the (un-located) value returned by the broadcast, attached with a proof that for all proofs |x| that can unwrap the original message |msg|, the value of |msg x| equals the resulting value |v|.
Intuitively, this means a message value does not change through broadcasting for roles that already had knowledge of the message.
By utilizing this attached proof, we are able to eliminate the dead branches found in MultiChor that we could previously not.
Note that we use variants |share| and |bcast| in this paper to drop the attached proofs of |share'| and |bcast'| in places where they are not needed.

\Cref{lst:nodead} shows the same protocol in ChorLean, where we can reuse evidence |h| of | ∀ p, (val p).isLeft = true| (line \ref{no-dead-24}) and |h'| from the first pattern match, to safely extract the correct side value from the sum (line \ref{no-dead-1}, line \ref{no-dead-2}).

\begin{lstlisting}[caption={No Redundant Pattern-Match on Sum in ChorLean}, label={lst:nodead}, numbers=left, float=b]
example {N t p}
  (val: (Sum LongList LargeMatrix) @[Frida] # t)
  : Choreo 2 t [Frida, Emil] p Unit := do
  match <- bcast' λ x => (val x).isLeft with
  | ⟨true, h⟩ =>/-@\label{no-dead-24}@-/
    let list := λ x =>
      (val x).getLeft (h x)/-@\label{no-dead-1}@-/
    handleList list
  | ⟨false, h'⟩ =>
    let matrix := λ x =>
      (val x).getRight (Sum.isLeft_eq_false.mp (h' x))/-@\label{no-dead-2}@-/
    handleMatrix matrix
\end{lstlisting}

A limitation of this approach is that communication still loses information of communicated message values at the receiving side. EPP of communication maps the receiver of a message to a receive operation, whose result depends on the real world network, and thus cannot be guaranteed to be the same value that has been sent.
In our example, |Frida| keeps knowledge of the broadcast value being constructed by either |inl| or |inr|, but this knowledge cannot be transferred to |Emil|.
Hence, properties that depend on correctly transferred messages cannot be proven by this system, such as termination of a distributed sorting algorithm.

ChorLean is able to reduce the amount of pattern matching needed in cases where message is not transferred to a different role.

%% file: chorlean.tex
\section{A Safe Choreographic Library with Dependent Types}
\label{s:implementation}

In this section, we present ChorLean, a library for dependently typed choreographic programming embedded in Lean.\footnote{While Lean is more well-known for theorem proving, it is also a fully featured dependently typed programming language, capable of compiling executable programs.}
In \Cref{ss:basic}, we introduce ChorLean’s design, 
following the state of the art in library-based approaches~\cite{ChoRus},
and explain how to program in ChorLean.
Then, in \Cref{ss:case-studies} we present what kind of case studies we were able to implement using ChorLean.
Finally, in \cref{ss:karatsuba} we discuss a particular case study in comparison to its MultiChor counterpart.

\subsection{ChorLean}
\label{ss:basic}

We present ChorLean's choreography and process types, and show how we carry proofs throughout endpoint projection to achieve totality.

\paragraph{Synchrony}
In choreographic libraries, the projected processes are designed to run in lockstep with respect to reaching the communication primitives.
When executed together, each role follows the same control flow through the choreography.
Because Lean and Haskell are pure languages, the same input deterministically determines the same branch of computation on every machine.
This synchrony ensures that communication actions are always matched by the intended sender and receiver, which is the library approach's key argument for deadlock freedom.

\paragraph{Enclaves and Located Values.}

In ChorLean, choreographies can include sub-choreographies (enclaves) that involve only a subset of roles within the overall choreography.
Library-based embeddings, unlike standalone languages, typically rely on broadcasting values to all participants before branching on these values is possible.
This ensures that all projections follow the same control flow through the program, because a variable contains the same value for all projections after broadcasting.
To reduce unnecessary messages, the concept of enclaves~\cite{ChoRus} was introduced. %
Even though one has to use broadcasting for control flow, restricting the target of sub-choreographies to a subset of roles results in said broadcasts only disseminated among those involved.

Enclave operations result in \emph{located values} that are only accessible (i.e., \emph{owned}) by those roles involved in the enclave.
The type of located variables ensures that processes never try to read variables they do not have knowledge of (see \Cref{s:total-epp}).
The target is a type parameter on the located type (see \Cref{ss:total}),
and inherited from the enclosed choreography, which has the target also as a type parameter.
A located variable can be read by choreography code only when every role currently executing the choreography (the census) is among the variable's owners. If even one role in the census does not own the variable, the value remains opaque.

\begin{lstlisting}[numbers=none, caption={Accessing Located Values in Choreographies.}, label={lst-located-access}]
def Alice: Fin 2 := 0
def Bob:   Fin 2 := 1
def bobValue {t}: String @ [Bob] # t := fun _ => "hello"/-@\label{lin:la-0}@-/

example {t p} : Choreo 2 t [Alice, Bob] p Unit := do
  -- Census = [Alice, Bob]
  -- Alice is not an owner, so the bobValue is opaque.
  let s := bobValue p    -- ERROR/-@\label{lin:la-1}@-/
  enclave [Bob] fun p' => /-@\label{lin:la-2}@-/
    -- Census = [Bob]
    let s := bobValue p' -- OK/-@\label{lin:la-3}@-/
    run (IO.println s)
\end{lstlisting}

\Cref{lst-located-access} illustrates this ownership discipline concretely.
In the example, |bobValue| (line \ref{lin:la-0}) is a located value owned exclusively by Bob.
At line \ref{lin:la-1}, Alice attempts to access it within a census of Alice and Bob.
This is rejected by the type checker, because the proof |p : t.v ∈ [Alice, Bob]| does not satisfy the stricter requirement |t.v ∈ [Bob]| that |bobValue| demands.
Access is only permitted inside an enclave restricted to Bob alone (line \ref{lin:la-2}), where the proof |p': t.v ∈ [Bob]| is available, like in line \ref{lin:la-3}.
This statically enforces that no role can observe a value it does not own, without any runtime checks.

\paragraph{Hidden Target Role.}
To guarantee totality for EPP, we added the projection target to choreographies and to located values as a parameter.
The inclusion of target |t| as a plain variable as shown in \Cref{s:total-epp} imposes a new problem, as it takes a different value for each projection.
If we had a choreography whose control flow depends on the value of |t|, we could easily break the invariant that all projections must follow the same control flow, effectively running different protocols on different roles.
To prevent this, user code is not allowed to inspect
the value of |t|.
To this end, we define the |Hidden| structure whose constructor and destructor are private, so that only our library can inspect target values:
\begin{lstlisting}
structure Hidden a where
  private mk ::
  private v: a
\end{lstlisting}
While hiding variables behind a safe API is undesirable, it currently is an inherent shortcoming of the Choreography-as-a-library approach.
The fact that private variables are transparent to choreographies written by the user is not exploitable by the Lean theorem proving mechanisms. 
This makes it impossible to prove directly in the theorem prover itself the usual guarantees expected by choreographies, such as deadlock freedom.
The HasChor, MultiChor and ChoRus approaches also rely on hiding via visibility modifiers.
ChorLean narrows down the usage of visibility modifiers to the bare minimum.
While other approaches hide the construction and destruction of \emph{all} located values (see \Cref{ss:total}), ChorLean gets away with only hiding the exact value of the projection target.
Previous definitions of located types also indirectly leaked information of the target to the user.
Through pattern matching, a user can check if a variable has been constructed as either |some| or |none|.
Our |Located| definition as a function does not leak this information, allowing us to expose it as is to the user, instead of having to provide a safe API like others.

\paragraph{Choreographies.}
In ChorLean, choreographies are represented by the |Choreo| type in \Cref{lst-choreo}.

As such, each choreography is parametrized by
(1)~the number of roles |N| which may appear in the choreography,
(2)~a target role |t| which is \emph{hidden} from the user whose value is supplied by library code,
(3)~the list of roles that participate in the protocol called |c|,
(4)~a proof |p| that |t.v| is a member of the census, and
(5)~the type \lstinline[language=lean]|α| that one will get by executing the choreography.
|Choreo| (\Cref{lst-choreo}) is an inductively defined type of choreographic primitives.
Note that we elide some proof terms that Lean's automated tactics can find, writing |∎| instead.

\begin{lstlisting}[numbers=left, caption={Choreo Monad.}, label={lst-choreo}, float=t]
inductive Choreo (N: Nat) (t: Hidden (Fin N)):/-@\label{lin:choreo-0}@-/
  (c: List (Fin N)) → 
  (p: t.v ∈ c)
  (α: Type) → Type 1
where
| Share {c os p α β} [Serialize β]/-@\label{lin:choreo-6}@-/
  (s r: (Fin N)) (ex: [s, r] ⊆ c) (msg: β @s::os #t)
  (next: {lv: β @r::s::os #t // ∀ x, msg x = lv ∎}/-@\label{lin:choreo-7}@-/
         → Choreo N t c p α)
  : Choreo N t c p α
| Run {c p α β} (r: (Fin N)) (alone: ∀ x ∈ c, x = r)/-@\label{lin:choreo-2}@-/
  (prog: IO β) (next: β → Choreo N t c p α)
  : Choreo N t c p α
| Enclave {c p α β } (c': List (Fin N)) (is_sub: c' ⊆ c)/-@\label{lin:choreo-3}@-/ 
  (schoreo: (h:t.v∈c') → Choreo N t c' h β)/-@\label{lin:choreo-4}@-/
  (next: (β @c' #t) → Choreo N t c p α)
  : Choreo N t c p α
| Return {c p α} (v: α) : Choreo N t c p α/-@\label{lin:choreo-ret}@-/
\end{lstlisting}

|Enclave| (line \ref{lin:choreo-3}) takes as arguments (1)~a sub-census |c'|, (2)~a proof |is_sub| that all roles in the new census are part of the previous census, (3)~a sub-choreography |schoreo: (h: t.v ∈ c') → Choreo N t c' h β|, which is provided a proof that the sub-choreography is only ever executed when the target role is part of the sub-census (allowing the sub-choreography to access values owned by the sub-census);
(4)~the remaining sequence of choreographic primitives is provided with the result of the sub-choreography restricted to that sub-census (|next : β @c' #t → Choreo N t c p α|).

|Share| (line \ref{lin:choreo-6}) takes as arguments (1)~a sender role |s| and (2)~a receiver role |r|, (3)~a proof |ex| that both roles are part of the census |c|, (4)~a list of owners |os|, (5)~a message |msg: β @s::os #t| which is located on the sender |s| and the owners |os|, and (6)~the remaining sequence of choreographic primitives called |next|.
During execution the message is to be shared with the receiver |r|, where the sharing is modeled by allowing the remaining operations to depend on a value of type |β @r::s::os #t|.

|Run| (line \ref{lin:choreo-2}) takes as arguments (1)~a role |r|, (2)~a proof |alone| that all roles in the census are equal to this role, (3)~an effectful IO program |prog|, and (4)~the remaining sequence of choreographic primitives called |next|, which may depend on the result of the program, e.g., a value of type |β|.

|Return| (line \ref{lin:choreo-ret}) takes as argument a single value |v| of type |α| and concludes the choreography.

\begin{lstlisting}[numbers=left, caption={Process Monad.}, label={lst:process}, firstnumber=21, float=t]
inductive Process {N: Nat} (t: Hidden (Fin N)):
  (α: Type) → Type 1
where
| Send {α β} [Serialize β]:/-@\label{lin:process-1}@-/
  (r: Fin N) →
  (p: t.v ≠ r) →
  (msg: β) →
  (next: Unit → Process t α) →
  Process t α
| Recv {α β} [Serialize β]:/-@\label{lin:process-2}@-/
  (s: Fin N) →
  (p: t.v ≠ s) →
  (next: β → Process t α) →
  Process t α
| Run {α β} (e: IO β)/-@\label{lin:process-3}@-/
  (next: β → Process t α):
  Process t α
| Return {α} (v: α): Process t α/-@\label{lin:process-4}@-/
\end{lstlisting}

\paragraph{Processes.}
The |Process| monad (\Cref{lst:process}) has four constructors.
The |Send| constructor represents sending a message over the network to a receiver |r| (line \ref{lin:process-1}),
while the |Recv| constructor receives a message from a sender |s| (line \ref{lin:process-2}).
The type of messages |α| has to implement the |Serialize| typeclass that allows for conversion from and into raw bytes for network transmission.
Both sending and receiving constructors demand inequality of the communication partner and |t|.
This makes sure a program does not unnecessarily exchange messages over the network interface with itself.
Further, |Run| (line \ref{lin:process-3}) performs an effectful IO program along with a continuation |next| of that effect's result, and
|Return| (line \ref{lin:process-4}) consists of only a single value |v| to be returned, concluding the process.

\begin{lstlisting}[numbers=left, label={lst:epp}, caption={Endpoint Projection}, firstnumber=41, float=t]
def Choreo.epp {N c α} (t: Hidden (Fin N)) {p}:
  Choreo N t c p α → Process t α
  | Choreo.Enclave c' _ subchoreo next =>
  if h: t.v ∈ c' then do
    let v <- epp t (subchoreo h)/-@\label{lin:epp-enclave-action}@-/
    epp t (next λ _ => v)
  else
    epp t (next λ h' => /-@\colorbox{goodybg}{False.elim (h h')}@-/)/-@\label{lin:epp-enclave-skip}@-/

  | Choreo.Share s r _ex msg next (os:=os) => do
    if h1: r ∈ s::os then/-@\label{lin:epp-com-skip}@-/
      epp t (next ⟨λ h' => msg ∎, ∎⟩)
    else if h2: t.v = s then /-@\label{lin:epp-send}@-/
      Process.Send r ∎ (msg ∎) λ () =>
        epp t (next ⟨λ h' => (msg ∎), ∎⟩)
    else if h3: t.v = r then/-@\label{lin:epp-recv}@-/
      Process.Recv s h2 λ v =>
        epp t (next ⟨λ _h' => v, ∎⟩)
    else if h4: t.v ∈ os then/-@\label{lin:epp-known}@-/
      epp t (next ⟨λ h' => msg ∎, ∎⟩)
    else/-@\label{lin:epp-nope}@-/
      epp t (next ⟨λ h' =>
      let h:t.v ∉  r::s::os := ∎
      /-@\colorbox{goodybg}{False.elim (h h')}@-/, ∎⟩)

  | Choreo.Run r alone prog next => do/-@\label{lin:epp-run}@-/
    Process.Run prog fun x => (next x).epp t

  | Choreo.Return v => Process.Return v/-@\label{lin:epp-return}@-/
\end{lstlisting}

\paragraph{Endpoint Projection.}
\Cref{lst:epp} shows EPP of |Choreo|.
In lines \ref{lin:epp-enclave-skip} and \ref{lin:epp-nope} we highlighted the impossible case in green.
We show |False| for every unreachable branch, and close it by |False.elim|, satisfying the totality check of Lean.
We abbreviated proofs in the listing by $\Box$, where the automatic proving tactics of Lean (|grind|) were sufficient to close the goal. 
The endpoint projection function |epp| translates choreographies to processes sharing the same target |t| and computation result type |α|.
Choreographies are a sequence of choreographic primitives, ending in a final |Return|. These primitives are translated recursively.
In particular, the communication primitive |Share| is translated depending on the target in one of five possibilities:
(1) if the receiver |r| of a message is identical to the sender |s| or a member of |os|, the message is already known to |r| (|r ∈ s::os|, line \ref{lin:epp-com-skip}), no process action is produced. This is an optimization which avoids unnecessarily retransmitting values over the network,
(2) if the target of the projection is the sender (|t.v = s|, line \ref{lin:epp-send}), a send operation is produced,
(3) if the target is the receiver (|t.v = r|, line \ref{lin:epp-recv}), a receive operation is produced,
(4) if the target is among the owners, EPP unwraps the located message (line \ref{lin:epp-known}) 
(5) otherwise the target is neither sender nor receiver nor owner,
    meaning this target is irrelevant to this communication operation and no process action is produced.
    In this case, the message was already inaccessible to the current target,
    and thus we do not need to provide a value for the continuation, and can instead make use of |False.elim| (line \ref{lin:epp-nope}).

|Choreo.Run| is always projected to a |Process.Run| constructor (line \ref{lin:epp-run}) that recursively projects the continuation.
The |Choreo.Enclave| primitive is translated to the translation of the enclosed choreography (line \ref{lin:epp-enclave-action}) if the target of the projection is part of the sub-census of the enclave operation, otherwise no process operation is produced (line \ref{lin:epp-enclave-skip}).
The case split in lines \ref{lin:epp-enclave-action}--\ref{lin:epp-enclave-skip} is required because the continuation |next| expects a value of type |β @ c' # t|, which is a function |t.v ∈ c' → β|. Such a function can only be produced when we have a proof of |t.v ∈ c'|; in the |else| branch this proof is obtained by contradiction (|False.elim|), and in the |then| branch the sub-choreography is projected to a value of type |β| which is then embedded as a constant function. The naive formulation |epp t (next (λ h => epp t (subchoreo h)))| is rejected by the type checker because |subchoreo| expects |h : t.v ∈ c'|, which is unavailable outside of the |then| branch.
In all cases, translation continues recursively with the continuation (|epp t (next ...)|) applied to the result of the previous translated operation.
The projection finishes by translating the final choreographic |Return| operation to a process |Return| operation (line \ref{lin:epp-return}).

As we described in \Cref{s:branches}, in the |next| continuation for communication (line \ref{lin:choreo-7} of \Cref{lst-choreo}), the message |v| carries a proof that the message value does not change for all owners |∀ x, msg x = v ...|.
EPP constructs the proofs for transported values accordingly.
Constructing the proofs has to distinguish between several cases:
\begin{enumerate}
  \item The receiving role |r| is either equal to |s| or part of |os|. In this case no network communication takes place, because the message value is known to |r| beforehand. The resulting value can be constructed from |msg ...|, and the proof can be constructed by reflexivity (line \ref{lin:epp-com-skip}).
  \item The target role is equal to |s|. Since |msg| is located at |s|, the resulting value can be mapped to the content of |msg| and the proof can be constructed by reflexivity (line \ref{lin:epp-send}).
  \item The target role is equal to |r|. In this case, the proof is constructed by false elimination, using evidence that the target must not already own the message (line \ref{lin:epp-recv}).
  \item Otherwise, the target role is neither the sender nor the receiver nor among the owners. This means the message is irrelevant to this projection. The proof of inaccessibility is established by contradiction.
\end{enumerate}

\paragraph*{Running Choreographies}

One possibility of running choreographies is the compilation into separate executable programs, one per possible target where
we have to make sure that programs enter the choreographic code in a coordinated way.
ChorLean allows compiling into a single executable program that all roles run.
\begin{lstlisting}[numbers=none, caption={Choreographic Main}, label={lst:cmain}]
def Faceted {N} t α :=  (r:(Fin N)) -> α @ [r] # t/-@\label{lin:main1}@-/

def ChoreoMain N := /-@\label{lin:main2}@-/
  {t: Hidden (Fin N)} ->
  Faceted t (List String) ->
  Choreo N t (List.finRange N) ∎ (Faceted t UInt32)/-@\label{lin:main5}@-/

def CHORLEAN_MAIN/-@\label{lin:main3}@-/
  (main: ChoreoMain N)
  (ips: Fin N → Fin N → IP_Address := defAddresses)/-@\label{lin:main6}@-/
  : List String → IO UInt32/-@\label{lin:main4}@-/
\end{lstlisting}
A user provides a choreographic main function |m| and a mapping from roles to physical network addresses |f|, and ChorLean generates a regular main function with |CHORLEAN_MAIN m f| (\cref{lst:cmain} lines \ref{lin:main3}-\ref{lin:main4}).
By default, ChorLean maps roles to local IPv4 addresses with the |defAddresses| function (line \ref{lin:main6}).
|CHORLEAN_MAIN| produces a main program that runs the projection which corresponds to the role of the first command line argument.
Subsequent command line parameters are passed to the choreographic main function.
Defining the |main| function of a Lean program in the shape |main := CHORLEAN_MAIN m f| offers developers a convenient way to implement endpoint programs for a choreography.

A choreographic main function is of type |ChoreoMain N| (lines \ref{lin:main2}-\ref{lin:main5}) and maps a target |t| and faceted command line arguments to a choreography that computes a |Faceted| Status code.
A |Faceted| value (line \ref{lin:main1}) maps a role to a value located at this role, in this case it maps a role to the command line arguments located at that role.

%% file: casestudies.tex
\subsection{Case Studies}
\label{ss:case-studies}

ChorLean's primitive operations suffice to compose complex choreographies.
We provide similar definitions to those in MultiChor from parallel computing paradigms, like broadcasting, parallel execution, scatter, gather, and allgather that are used in our case studies.
Our artifact~\cite{daniel_2026_21531312} provides implementations of several case studies that have been previously presented by other choreographic libraries and languages, to confirm ChorLean's practicality for real-world distributed protocols and how our system applies to them.
Namely we implemented the authentication protocol from the Choral paper~(\cite{Choral} 3.1 Distributed authentication), the Goldreich-Micali-Wigderson protocol from the MultiChor paper~\cite{MultiChor}, and the distributed Karatsuba multiplication algorithm found in Choral, HasChor~\cite{HasChor} and MultiChor.

  The authentication protocol coordinates credential validation across Client, Service, and Identity Provider roles using salted hashes and tokens.
  The Goldreich-Micali-Wigderson (GMW) protocol implements secure multiparty computation over boolean circuits, using secret sharing and oblivious transfer.
  The Karatsuba multiplication distributes integer multiplication across multiple roles.

The Karatsuba multiplication algorithm demonstrates a compute-focused concurrent scenario.
As an example, we describe the Karatsuba multiplication case study in more detail and discuss the implementation differences to the MultiChor variant.

\subsection{Karatsuba Multiplication}
\label{ss:karatsuba}

The Karatsuba multiplication choreography distributes integer
multiplication across three roles |a|, |b| and |c| (\Cref{lst:karatsuba}, line~\ref{lin:karatsuba-def}).
Located inputs |n1| and |n2| are owned by role |a|.

The algorithm begins by broadcasting a boolean termination condition |g| (line~\ref{lin:karatsuba-done}).
If either input is less than 10, the product is computed directly and returned as a
located value (line~\ref{lin:karatsuba-early-return}). Otherwise, the helper function |f| splits each
input into high and low halves at the midpoint (lines~\ref{lin:karatsuba-f1} - \ref{lin:karatsuba-f2}), and the
sub-products |z0|, |z1|, |z2| are computed via three recursive calls
distributed across the roles (lines~\ref{lin:karatsuba-rec1}-\ref{lin:karatsuba-rec2}). The final result is computed at
role |a| in lines~\ref{lin:karatsuba-return1}-\ref{lin:karatsuba-return2}.

The fact that the algorithm is marked as a partial definition does not conflict with our established totality guarantees for located types, instead it marks this function as a possibly infinite loop.
Termination of the recursion depends on the inputs decreasing at each step, which in a distributed setting additionally assumes correct message delivery.
For a provably terminating definition, the algorithm could instead fall back to a bounded implementation, which terminates after a fixed number of steps.

The main entry point (line~\ref{lin:karatsuba-main}) reads command-line arguments only for
role~0 and wraps them in an |Option| located value (lines~\ref{lin:karatsuba-input1}-\ref{lin:karatsuba-input2}).
A boolean guard is broadcast via |bcast'| (line~\ref{lin:karatsuba-guard}) to synchronize all
roles on input validity. In the |true| branch, Lean's |grind|
tactic automatically discharges the proof obligation needed to apply to
|Option.get| (line~\ref{lin:karatsuba-getopt}), eliminating the redundant |none| case
that would be required in a non-dependently typed setting.

\begin{lstlisting}[numbers=left, label={lst:karatsuba}, caption={Karatsuba parallel Multiplication algorithm in ChorLean}, float]
structure KaratsubaNums where
  splitter: Int
  h1: Int
  h2: Int
  l1: Int
  l2: Int

def f (n1 n2:Int): KaratsubaNums :=/-@\label{lin:karatsuba-f1}@-/
  let m := max (log10 n1) (log10 n2) + 1
  let m2 := ((m / 2).floor).toInt64.toInt
  let splitter := (10:Int) ^ m2.toNat
  let h1 := n1 / splitter
  let l1 := n1 %
  let h2 := n2 / splitter
  let l2 := n2 %
  ⟨splitter, h1, h2, l1, l2⟩/-@\label{lin:karatsuba-f2}@-/

partial def karatsuba {N t census p}/-@\label{lin:karatsuba-def}@-/
  (a b c: Fin N)
  (n1 n2: Int @ [a] # t)
  (_: [a,b,c] ⊆ census)
  : Choreo N t census p (Int @ [a] # t) := do

  let g <- bcast λ h =>  n1 h < 10 || n2 h < 10/-@\label{lin:karatsuba-done}@-/
  match g with
  | true =>
    return λ h => (n1 h) * (n2 h)/-@\label{lin:karatsuba-early-return}@-/
  | false =>
    let x   := λ h => f (n1 h) (n2 h)/-@\label{lin:karatsuba-comparison}@-/
    let l1' <- com b λ h => (x h).l1
    let l2' <- com b λ h => (x h).l2
    let h1' <- com c λ h => (x h).h1
    let h2' <- com c λ h => (x h).h2
    let z0 <- karatsuba b c a l1' l2' ∎ >>= com a/-@\label{lin:karatsuba-rec1}@-/
    let z2 <- karatsuba c a b h1' h2' ∎ >>= com a
    let s1 := λ h => (x h).l1 + (x h).h1
    let s2 := λ h => (x h).l2 + (x h).h2
    let z1' <- karatsuba a b c s1 s2 ∎/-@\label{lin:karatsuba-rec2}@-/
    let z1 := λ h => (z1' h) - (z2 h) - (z0 h)
    return λ h =>/-@\label{lin:karatsuba-return1}@-/
      let s := (x h).splitter
      (z2 h * s * s) + (z1 h * s) + (z0 h)/-@\label{lin:karatsuba-return2}@-/

def main := CHORLEAN_MAIN 3 λ args => do/-@\label{lin:karatsuba-main}@-/
  let z := args 0 -- command line args of role 0
  let input:(Option (Int × Int)) @ _ #_ := λ h =>/-@\label{lin:karatsuba-input1}@-/
    if h:(z h).length = 2 then
      some ⟨←((z h)[0]).toInt?, ←((z h)[1]).toInt?⟩
    else
      none/-@\label{lin:karatsuba-input2}@-/
  let guard <- bcast' λ h => (input h).isSome/-@\label{lin:karatsuba-guard}@-/
  match h:guard.val with
  | false => return λ _ _ => -1
  | true =>
    let v := λ h => (input h).get ∎/-@\label{lin:karatsuba-getopt}@-/
    let res <- karatsuba 0 1 2 
        (λ h => (v h).fst) (λ h => (v h).snd) ∎
    locally 0 λ h => IO.println s!"result: {res ∎}"
    return λ _ _ => 0
\end{lstlisting}

The MultiChor implementation of Karatsuba follows the same structure, but several differences highlight the advantages of ChorLean's dependent type system.
\begin{lstlisting}[numbers=left, label={lst:karatsuba-multi}, caption={Karatsuba Main from MultiChor}]
mainChoreo :: Choreo Participants (CLI IO) ()
mainChoreo = do
  n1 <- _locally primary $ getInput "First number:"/-@\label{lin:karatsuba-multi1}@-/
  n2 <- _locally primary $ getInput "Second number:"/-@\label{lin:karatsuba-multi2}@-/
  result <- karatsuba primary worker1 worker2 n1 n2
  primary `locally_` \un -> do
    putOutput "Result:" (un primary result)
\end{lstlisting}
In MultiChor, the choreographic main function (\Cref{lst:karatsuba-multi}) reads the two inputs via separate calls to |_locally| (lines~\ref{lin:karatsuba-multi1}-\ref{lin:karatsuba-multi2}), where the |getInput| function might fail and throws an error.
Without the guarantee that both inputs are present before the algorithm proceeds, the program might end up in a deadlock.
In the ChorLean version, the inputs are wrapped in an |Option| and a boolean guard is broadcast via |bcast'| to synchronize all roles on input validity.
Crucially, in the true branch Lean's grind tactic automatically discharges the proof that the |Option| is |.some|, so |Option.get| can be called without a redundant |.none| branch.
This branch would be required in MultiChor's Haskell setting, if you were to synchronize on input validity safely, or else the complete input would need to be sent.

As discussed in \Cref{s:branches}, MultiChor requires undefined placeholders when pattern-matching on sum types after a broadcast, because the type system cannot track which constructor was chosen. ChorLean's proof-carrying located values eliminate this. The broadcast result of |bcast'| carries a proof that the message did not change its value, which eliminates the dead branch.

A further difference is how proof obligations for accessing located values are handled:
In MultiChor, a custom proof system for subset relations of roles is axiomatized, whose proofs are constructed manually.
E.g. in their Karatsuba code:
\begin{lstlisting}[numbers=none, label={lst:karatsuba-multi1}]
a `locally` \un ->
  return $ un singleton n1 * un singleton n2
\end{lstlisting}
corresponds to line \ref{lin:karatsuba-comparison} in the ChorLean version, where \newline |singleton n1| constructs a membership proof that |a| is a member of the census |[a]|.
In ChorLean, these membership proofs are just standard Lean obligations, so Lean tactics like grind can be used and we do not require additional axioms on top of Lean's builtin logic.
For all applications from case studies found in MultiChor, Lean was able to find the proof obligations to unwrap located values automatically, easing the proof burden on developers.

%% file: related-work.tex
\section{Related Work}
\label{s:related-work}

Classic choreographic approaches usually are implemented as \emph{standalone languages} such as AIOCJ~\cite{AIOCJ}, Choral~\cite{Choral}, and Kalas~\cite{Kalas} that compile a choreography into a set of endpoint programs, one per role.
In contrast, many recent approaches follow a \emph{library-based} implementation style, such as ChoRus~\cite{ChoRus}, HasChor~\cite{HasChor}, and MultiChor~\cite{MultiChor}, and embed choreographies in general-purpose languages like Haskell or Rust, performing projection at run time.
To the best of our knowledge, all existing library approaches are implemented using partial functions for accessing located values, in discrepancy with the formal theory. We address this using dependent types, making located value access total.

HasChor \cite{HasChor}, introduced by Shen et al.\@, was among the first libraries to leverage Haskell's strong type system for embedding choreographies. The primary mechanism for coordinating choices in HasChor is broadcasting, and the library does not support enclaves. As a consequence, all values required for protocol-level decisions -- such as branching -- must be sent to every role in the system, potentially resulting in unnecessary network communication.
HasChor implements located types akin to our approach in \Cref{ss:total},
but internally implements access to them with a partial function.
Also, in contrast to MultiChor, located values are only located at a single role.
Compared to HasChor, we support enclaves, values that are located at multiple roles, use dependent types to prove that located value access is total.
ChoRus \cite{ChoRus} extends the choreography-as-a-library approach to the Rust programming language. It improves upon HasChor by introducing an enclave mechanism (referred to as \emph{conclave}), in addition to broadcasting. This feature enables the execution of sub-protocols, allowing roles to avoid receiving broadcasted values when they are not required, thus supporting more efficient conditional constructs.
MultiChor \cite{MultiChor} is another Haskell library based on HasChor that introduces both \emph{conclaves} and values located at multiple roles, rather than a single role.
Accessing a located value requires a proof that the accessing role is part of the set of roles that have knowledge of the value. In MultiChor, this is achieved with a specialized proof
library, while ChorLean uses Lean's standard proving capabilities.
Further, MultiChor's implementation also uses a partial function for located value access.
Their \emph{conclave} primitive is not needed in ChorLean, as it is a side effect from hiding the located type API completely from the user, instead of just the target.

Shen et al.~\cite{Shen2024} present an initial sketch of how choreographic library implemented in the Agda proof assistant could use dependent types to enforce total access to located variables. To accomplish this, they parameterize choreographies over a function that maps located types to ordinary host language types. 
However, they provide no examples of executable choreographies that demonstrate the viability of this approach in practice.
In attempting to replicate their approach in Lean, we encountered several universe-level type errors, which led us to fall back on the |Hidden| wrapper for the projection target as described in \cref{ss:basic}.
Exploring alternatives that do not rely on visibility modifiers to ensure correctness remains an open direction for future work.

Multitier languages share similarities with choreographic programming but differ significantly in terminology and design \cite{multiparty}. In multitier languages, the entities participating in the system are typically referred to as \emph{tiers}. Unlike choreographies, which adopt a perspective from ``outside'' the roles, expressions in multitier programs are written from the viewpoint of a specific tier, using lexical scopes to nest expressions of different tiers.

Session types \cite{10.1007/3-540-57208-2_35} constitute a related research area with many conceptual parallels to choreographic programming. The central idea is to assign a global type to a protocol, describing the interactions between roles. These global types can then be projected onto individual roles,
enabling verification that local programs conform to the overall protocol. In contrast, choreographic programming works by projecting \emph{programs} onto individual roles.

%% file: conclusion.tex
\section{Conclusion}
\label{s:conclusion}

In this paper, we presented our implementation of a choreographic library in the dependently typed language Lean.

While existing choreographic libraries
use partial access operations for located values, we leveraged Lean's dependent type system
to make access operations total, which forced our implementation to handle them correctly.
This allows defining endpoint projection as a total function, as ensured by Lean's totality checker. We also used Lean's type system to avoid dead branches in user code which arise in prior approaches.

%% file: acknowledgments.tex
\begin{acks}
We thank the anonymous reviewers for their extensive and helpful feedback.
This research work was supported by the National Research Center for Applied Cybersecurity ATHENE. ATHENE is funded jointly by the German Federal Ministry of Education and Research and the Hessian Ministry of Higher Education, Research and the Arts.
This work was funded by the LOEWE initiative (Hesse, Germany)\\ \ [LOEWE/4a//519/05/00.002(0013)/95],
by the LOEWE initiative (Hesse, Germany) within the
emergenCITY center [LOEWE/1/12/519/03/05.001(0016)/72], and by the Deutsche Forschungsgemeinschaft (DFG, German Research Foundation) as part of Germany's Excellence Strategy –- EXC-3057/1 ``Reasonable Artificial Intelligence'' –- Project No. 533677015.
\end{acks} 
\newpage